\documentclass{aa}
\usepackage{graphicx}
\begin{document}
   \title{An estimate of the time variation of the O/H radial gradient from 
   planetary nebulae}

%   \subtitle{I. Overviewing the $\kappa$-mechanism}

   \author{W.J. Maciel, R.D.D. Costa, \and M.M.M. Uchida
          \inst{ }}

   \offprints{W.J. Maciel}

   \institute{IAG/USP, S\~ao Paulo, Brazil\\
              \email{maciel@astro.iag.usp.br, roberto@astro.iag.usp.br,
               monica@astro.iag.usp.br}}

   \date{Received ; accepted }

   \abstract{
   Radial abundance gradients are a common feature of spiral galaxies,
   and in the case of the Galaxy both the magnitude of the gradients 
   and their variations are among the most important constraints of 
   chemical evolution models. Planetary nebulae (PN) are particularly
   interesting objects to study the gradients and their variations.
   Owing to their bright emission spectra, they can be observed
   even at large galactocentric distances, and the derived abundances
   are relatively accurate, with uncertainties of about 0.1 to 
   0.2 dex, particularly for the elements that are
   not synthesized in their progenitor stars. On the other hand,
   as the offspring of intermediate mass stars, with main sequence 
   masses in the interval of 1 to 8 solar masses, they are representative
   of objects with a reasonable age span. In this paper, we present
   an estimate of the time variation of the O/H radial gradient in a 
   sample containing over 200 nebulae with accurate abundances. Our 
   results are consistent with a flattening of the O/H gradient 
   roughly from $-0.11$ dex/kpc to $-0.06$ dex/kpc during the last 9 Gyr, 
   or from $-0.08$ dex/kpc to $-0.06$ dex/kpc during the last 5 Gyr.
   }
  \authorrunning{Maciel, Costa and Uchida}
  \titlerunning{Time variation of the O/H gradient from PN}

   \maketitle
%
%________________________________________________________________

\section{Introduction}

Radial abundance gradients in the Galaxy have been determined
by a variety of objects, including  HII regions, planetary
nebulae (PN), B stars, open clusters and cepheids
(Henry \& Worthey \cite{henry}, Maciel \cite{maciel00},
Rolleston et al. \cite{rolleston}). Several elements
have been investigated, especially oxygen, sulphur, neon and
argon in photoionized nebulae (Maciel \& Quireza \cite{mq},
Deharveng et al. \cite{deharveng}) and iron in open cluster 
stars and cepheids (Friel \cite{friel}, Andrievsky et al.
\cite{sergueia}, \cite{sergueib}). In the case of the O/H ratio, recent 
determinations from HII regions, PN and B stars point to an 
average gradient in the range $-0.04$ to $-0.07$ dex/kpc, 
taking into account an average uncertainty of $0.02$ dex/kpc 
for most determinations.

Abundance gradients play a distinctive role as a 
constraint to chemical evolution models. In fact, several 
recently computed models, based on widely differing 
assumptions, point to the radial gradients as one of the most 
important constraints, especially when one takes into account 
not only their magnitudes, but also their space and time 
variations (see for example Hou et al. \cite{hou}, Chiappini
et al. \cite{cris}, and Alib\'es et al. \cite{alibes}).

Planetary nebulae are particularly useful in the study 
of the gradients and their variations (Peimbert 
\cite{mp90}, Maciel \& K\"oppen \cite{mk}, Maciel 
\cite{maciel97}, \cite{maciel00}, Peimbert \&
Carigi \cite{pc98}, Maciel \& Quireza \cite{mq}, 
Maciel \& Costa \cite{mcosta}). 
They usually have bright emission spectra, so that they can 
be observed even at large heliocentric distances. Their 
derived abundances are relatively accurate,  
with uncertainties of about 0.1 to 0.2 dex, especially 
regarding the elements that are not synthesized by their 
progenitor stars, such as neon, argon and to some extent
oxygen. On the other hand, as the offspring of intermediate 
mass stars, with main sequence masses roughly in the 
interval $1 < M/M_\odot < 8$, they are representative of 
objects spanning a relatively large age interval, so that 
it is expected that groups of PN of different ages may 
display different abundance variations, thus reflecting 
the time evolution of the gradients. The PN distances are 
a possible source of error, as they are not as well 
determined as in the case of B stars or HII regions, for
example. However, it has been shown that the use of 
different distance scales, both individual and statistical, 
apparently compensates for this uncertainty (see for
example Maciel \& K\"oppen \cite{mk}). Moreover, it should
be recalled that in order to determine the PN gradients 
larger samples are generally used as compared with 
HII regions.

In this paper, we present an estimate of the time variation
of the O/H radial abundance gradient based on a large sample 
of galactic PN with relatively accurate abundances and 
distances. From the observed oxygen abundances, we determine
the [Fe/H] metallicity using a correlation based on disk stars, 
and the progenitor ages are obtained through an age-metallicity 
relation. In section 2 we describe our sample, in section 3
we discuss the [O/H] $\times$ [Fe/H] relation for
disk stars. The ages of the objects are estimated
in section 4, and we present our method to determine 
the abundance gradients as a function of age. 
Finally, Section 5 presents our results and discussion.

%________________________________________________________________

\section{The data}

The sample includes about 240 nebulae, most of which 
have oxygen abundances by number $\log{\rm(O/H)}$
and distances from the samples of Maciel \& Quireza 
(\cite{mq}) and Maciel \& K\"oppen (\cite{mk}), 
to which the reader is referred for details 
on the abundances and references. About 40 new 
nebulae have been included, basically from
recent observations from our own group, secured at
the 1.52m ESO telescope at La Silla and the 1.60m LNA
telescope in Brazil (Costa et al. \cite{ccmf96}, 
\cite{ccmf97}, \cite{cmu02}). Most of the new objects 
belong to a project to derive accurate and homogeneous 
abundances of PN located near the anticentre direction. 
A detailed analysis of the new observational data, 
plasma diagnostics and abundances, as well as a study 
of the space variations of the abundance gradients 
along the galactic disk is given elsewhere 
(Costa et al. \cite{cmu02}). It should be mentioned 
that most O/H abundance determinations are based 
on the so-called \lq\lq empirical\rq\rq \ method, 
according to which the total abundances are obtained 
from ionic abundances with the use of ionization
correction factors for the unseen species. The ionic 
abundances depend on the plasma diagnostic parameters, 
namely the electron temperature and density, which are 
derived from selected line ratios (see for example
Peimbert \cite{mp90} and references therein). This 
procedure implies average uncertainties of about 
0.1 to 0.2 dex for the abundances of most elements
and up to 0.02 dex/kpc for the derived gradients.
These uncertainties are able to accomodate any
systematic variations with the galactocentric distance
(see for example Martins \& Viegas \cite{lucimara}).
In fact, the presence of abundance gradients can also 
be detected by considering abundances derived by other 
methods, such as photoionization models or detailed 
individual analysis, and the main discussions refer 
to the magnitude of the gradients.

Our sample consists only of disk nebulae, so that
PN of Types~IV (halo objects) according to the
Peimbert (\cite{p78}) classification system, or
of Type~V (bulge objects, see Maciel \cite{maciel89})
are not included. Most of the objects are classified
as Type~II  (disk objects, intermediate mass 
progenitors), but PN of Type I (disk objects, large mass
progenitors) and III (thick disk objects, showing 
large deviations from the galactic rotation curve) 
are also included, so that the average dispersion of the
oxygen abundances of the whole sample is expected
to be larger than in Maciel \& Quireza (\cite{mq}).

There are some suggestions that Type~I PN may show
some effects of ON cycling in their oxygen abundances,
but current evidences are not conclusive (Peimbert 
\& Carigi \cite{pc98}, Torres-Peimbert \& Peimbert 
\cite{silvia}). Furthermore, the amount of oxygen 
depletion from this process is expected to be much 
lower than the average abundance dispersion and 
restricted to the objects of larger masses, 
which make up a small fraction of the sample.

%________________________________________________________________

\section{The [O/H] $\times$ [Fe/H] relation}

The determination of the radial O/H abundance gradient
was performed in a similar way as in Maciel \& K\"oppen
(\cite{mk}), and simple linear fits have been obtained.
The O/H abundances by number of atoms have been converted 
into the usual [O/H] abundances relative to the Sun
using the solar oxygen abundance 
$\log{\rm (O/H)}_\odot + 12 = 8.83$ (Grevesse \& Sauval 
\cite{grevesse}), so that we have the relation

   \begin{equation}%eq.(1)
     {\rm [O/H]} = \log{\rm (O/H)}  - \log{\rm (O/H)}_\odot \ .
    \label{defoh}
   \end{equation}

\begin{figure*}
   \centering
   \includegraphics[angle=-90,width=13cm]{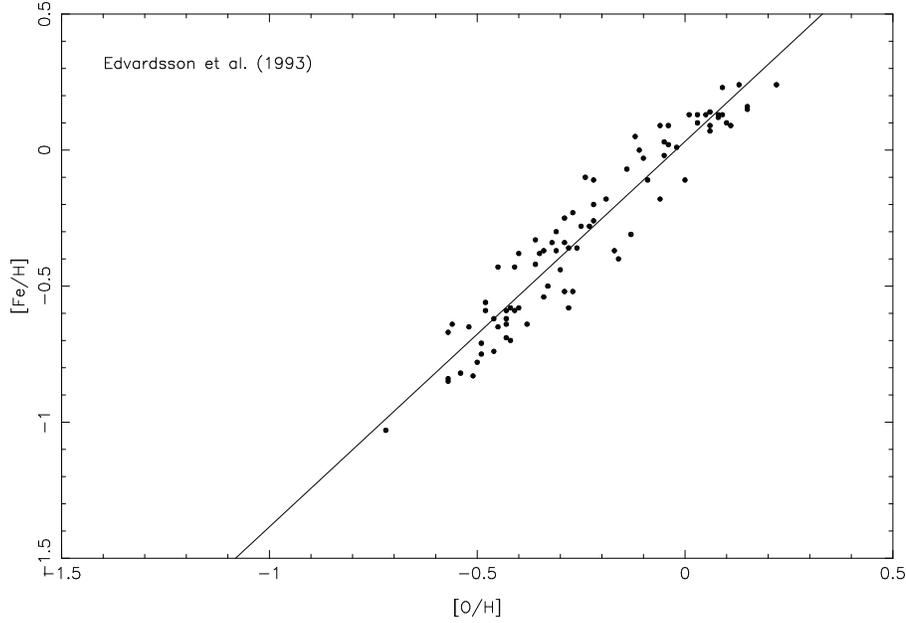}
      \caption{Correlation between [Fe/H] and 
      [O/H] obtained from the data of 
      Edvardsson et al. (1993).
              }
         \label{fig1}
   \end{figure*}

Iron abundances are very difficult to determine from
PN and their central stars, basically because the iron 
lines are very weak in PN spectra, and a presumably 
large fraction of iron is condensed in grains
(see for example Perinotto et al. \cite{perinotto}
and Deetjen et al. \cite{deetjen}). In order to convert 
from [O/H] to the iron metallicities [Fe/H], we have used 
the extensive sample of disk stars by Edvardsson et al.
(\cite{edvardsson}), from which we derive the
relationship

   \begin{equation}%eq.(2)
     {\rm [Fe/H]} = 0.0317 + 1.4168 \ {\rm [O/H]} \ ,
     \label{fehoh}
   \end{equation}

\noindent
valid approximately in the range
$-1.0<$ [Fe/H] $<0.5$. The average
uncertainty of the [Fe/H] metallicity
is roughly 0.1~dex. For this relation we have 
determined an average uncertainty for the
slope of 0.0488, and for the intercept
the uncertainty is 0.0158. The correlation
coefficient is $r = 0.9536$, which reflects
the good fit of Eq.~\ref{fehoh}, as can be
seen in Fig.~\ref{fig1}.

 \begin{figure*}
   \centering
   \includegraphics[angle=-90,width=13cm]{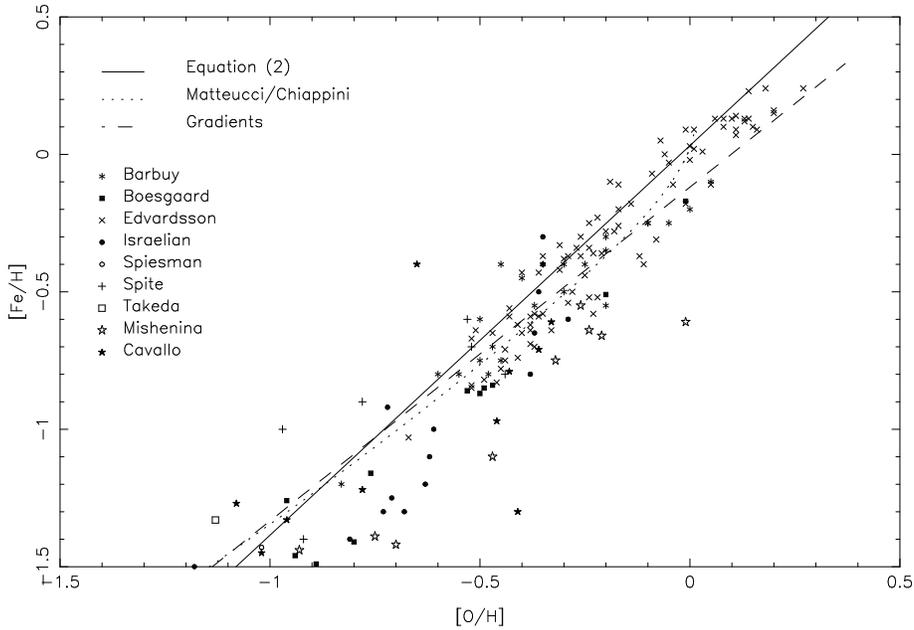}
      \caption{A comparison of the average relation given
       by Eq.~\ref{fehoh} (solid line), theoretical models by
       Matteucci et al. (1999, dotted line), the predicted
       relation derived from the gradients of young objects
       (Maciel 2002, dashed line) and selected
       observational data from a number of sources (see text).
              }
         \label{fig2}
   \end{figure*}

In order to test the applicability of Eq.~\ref{fehoh}
to the galactic disk as well as to the solar neighbourhood,
we have made some comparison with the [Fe/H] $\times$ [O/H]
relation as obtained for different sets of galactic objects
by different authors. A convenient set has been recently 
analysed by Maciel (\cite{maciel02}), in a study of the use 
of radial abundance gradients mainly derived from young 
objects in the determination of the [O/Fe] $\times$ 
[Fe/H] relation in the galactic disk. The main results 
are shown in Fig.~\ref{fig2}. In this figure, the solid 
line shows the relation given by Eq.~\ref{fehoh}. 
The dotted curve shows results of theoretical models by 
Matteucci et al. (\cite{matteucci99}), which follow models by 
Chiappini et al. (\cite{cris97}), and are representative of 
models predicting a [O/Fe] plateau for metallicities under 
solar, without a significant increase in the [O/Fe] ratio above
0.5 dex for [Fe/H] $\leq -2$. The dashed line is the 
predicted relationship on the basis of the radial gradients
of young objects in the galactic disk, at galactocentric
distances $4 \leq R ({\rm kpc}) \leq 12$, as discussed
by Maciel (\cite{maciel02}). The figure also includes some 
representative observational data by Barbuy \& Erdelyi-Mendes 
(\cite{barbuy}), asterisks; Boesgaard et al. (\cite{boesgaard}), 
filled squares; Edvardsson et al. (\cite{edvardsson}), crosses; 
Israelian et al. (\cite{israelian}), solid dots; Spiesman \& 
Wallerstein (\cite{spiesman}), open circles; 
Spite \& Spite (\cite{spite}), plus signs; Takeda et al. 
(\cite{takeda}), empty squares; Mishenina et al. (\cite{mishenina}), 
open stars, and Cavallo et al. (\cite{cavallo}), filled stars, 
as recomputed by Mishenina et al. (\cite{mishenina}).

It can be seen from Fig.~\ref{fig2} that the observational data 
and models show a reasonable agreement, especially for 
metallicities close to and slightly lower than the solar value.  
For the lowest metallicities the spread is somewhat larger
and the points lie, on the average, under the line
corresponding to Eq.~\ref{fehoh}, but the relatively sparse 
data are still reasonably well represented by the solid line. 
Part of the scatter in Fig.~\ref{fig2} may be due to the use of 
different scales of stellar parameters such as effective temperatures, 
gravities and metallicities, to the adoption of different atomic 
parameters or the neglecting of NLTE effects. The relation 
given by the theoretical models (dotted line) is representative 
of the galactic disk, since the calculated variations of the 
[O/Fe] $\times$ [Fe/H] relation from galactocentric distances 
$R = 4\,$kpc to $R = 14\,$kpc are expected to be small
(Matteucci \cite{matteucci96}). The gradient derived line (dashed 
line) was based on O/H and [Fe/H] gradients for young objects 
(HII regions, hot stars and open clusters) located at galactocentric 
radii roughly in the range $R = 4\,$kpc to $R = 16\,$kpc. On the 
other hand, the stellar data include objects in the thin disk, 
thick disk and even some metal-poor halo stars. Therefore, it can 
be concluded that Eq.~\ref{fehoh} represents fairly well the average 
\hbox{[Fe/H] $\times$ [O/H]} relation in the disk, so that any 
galactocentric variation of this relation is absorbed by the 
expected scatter, as shown in Fig.~\ref{fig2}. 

%________________________________________________________________

\section{The [Fe/H] $\times$ $\tau$ relation}

The main difficulty in the estimate of the time variation
of the abundance gradients from planetary nebulae lies
in the determination of reliable ages for the central stars.
One possibility is to use an average age-metallicity 
relation, since chemical abundances from PN are relatively
well determined, and several age-metallicity relations
derived recently are similar, albeit with a considerable
scatter (see Rocha-Pinto et al. \cite{rp2000} for a 
recent discussion). In fact, there has been some discussion
on the existence of such a relationship (Feltzing et al.
\cite{feltzing}), but a critical analysis of the available
data suggests some increase in the average metallicities
with time, which is expected on the basis of the current
ideas on the chemical evolution of the Galaxy.

Age-metallicity relationships have been determined by a
number of people (Twarog \cite{twarog}, Carlberg et al.
\cite{carlberg}, Meusinger et al. \cite{meusinger},
Edvardsson et al. \cite{edvardsson}, Rocha-Pinto \&
Maciel \cite{rpm98}, Rocha-Pinto et al. \cite{rp2000}),
based on a variety of samples and techniques. In general,
these relationships are similar, which is particularly 
true for the relations derived by Edvardsson et al.
({\cite{edvardsson}) and the recent results based on
chromospheric ages by Rocha-Pinto et al. 
(\cite{rp2000}). This can be seen from Fig. 14 of
Rocha-Pinto et al. (\cite{rp2000}), where the new
relation is compared with the relation by Edvardsson
et al. (\cite{edvardsson}) adopting 1 Gyr average bins,
as well as with the results by Twarog (\cite{twarog}),
Carlberg et al. (\cite{carlberg}) and Meusinger et
al. (\cite{meusinger}). These results apply strictly
to the solar neighbourhood, since most of the stars
included in the samples are nearby objects with 
HIPPARCOS parallaxes. However, the observed scatter
in these relationships is considerably large, amounting
up to 0.26 dex for the results of Edvardsson et al.
(\cite{edvardsson}) and about 0.13 dex for the relation
obtained by Rocha-Pinto et al. (\cite{rp2000}), so that
it probably includes any differences in the corresponding
relationships at different galactocentric distances.
This can be seen, for example, in Fig. 14a of Edvardsson
et al. (\cite{edvardsson}), where the metallicity [Fe/H]
is plotted against age using different symbols for stars
at galactocentric distances $R < 7\,$kpc, 
$7 < R ({\rm kpc}) < 9$ and $R > 9\,$kpc, spanning
radii from 4 kpc to 11 kpc (see also Table 14 of
Edvardsson et al. \cite{edvardsson}). Even though
the sample at $7 < R ({\rm kpc}) < 9$ is the most
complete statistically, it can be seen that all objects
can be reasonably fit in the average age-metallicity
relation, in view of the large scatter. 

In order to take into account any difference 
in the age-metallicity relationship 
introduced by differente galactocentric
distances, Edvardsson et al. (\cite{edvardsson})
went a step further and derived an average
age-metallicity-radius relation given by

   \begin{equation}%eq.(3)
     \log \tau = 0.872 - 0.303\ {\rm [Fe/H]} - 0.038\ R \ ,
     \label{taufeh}
   \end{equation}

\noindent
where $\tau$ is the stellar age in Gyr and
$R$ is the galactocentric distance in kpc
(see eq. 19 of Edvardsson et al. \cite{edvardsson}),
with a typical uncertainty of $\sigma(\log \tau) \simeq 0.10$.
This equation was based on stars with 
$\log\tau < 1.0$, a condition fulfilled by all
objects in our sample, the majority of
which have $\tau \simeq 5\,$Gyr, or 
$\log \tau \simeq 0.70.$

 \begin{figure}
   \centering
   \includegraphics[angle=-90, width=8cm]{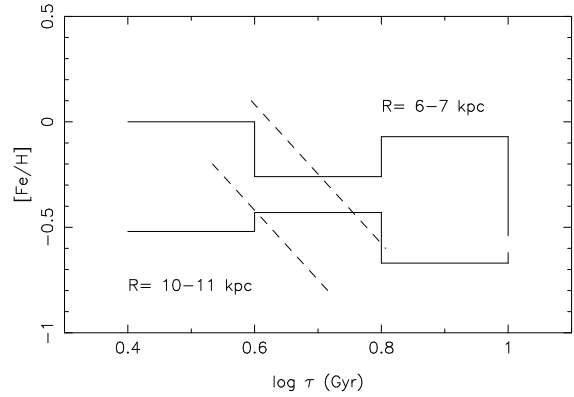}
      \caption{A comparison of age-metallicity relation as
       derived by Edvardsson et al. (\cite{edvardsson}) for
       stars in the galactocentric ranges $6 < R ({\rm kpc}) < 7$
       (top histogram) and $10 < R ({\rm kpc}) < 11$
       (bottom histogram) with the estimated relation
       calculated by Eq.~\ref{taufeh} (dashed lines),
       where we have used $R = 6.5\,$kpc and 
       $R = 10.5\,$kpc, respectively.
        }
         \label{fig3}
   \end{figure}

We can compare the ages derived by Eq.~\ref{taufeh} with the actual
average age-metallicity relation by Edvardsson et al.
(\cite{edvardsson}) for different galactocentric distances
as given in Table 14 of that paper. We performed this
calculation for the galactocentric radii 
$4 < R ({\rm kpc}) < 11$, and Fig.~\ref{fig3} shows 
representative examples for $6 < R ({\rm kpc}) < 7$ and
$10 < R ({\rm kpc}) < 11$. The histogram data
are from Table 14 of Edvardsson et al. (\cite{edvardsson})
while the dashed lines are results from the application
of Eq.~\ref{taufeh}, adopting $R = 6.5\,$kpc and 
$R = 10.5\,$kpc, respectively. It can be seen that the 
dashed lines produce similar results as the age-metallicity 
relation, considering that our objects have ages $\log \tau < 1\,$Gyr,
peaking at $\tau \simeq 0.7\,$Gyr, as already mentioned.
Therefore, the use of Eq.~\ref{taufeh} introduces a correction
to the age-metallicity relation for the solar neighbourhood
adjusting it to other galactocentric distances in the
range of 4 to 11 kpc approximately. Of course, the 
uncertainties in the derived ages are still considerably
large, irrespective of the galactocentric radius,
as can be seen by the scatter in the age-metallicity 
relations quoted here. However, in this
study we are mainly interested in the time {\it variation}
of the gradients, so that, in fact, only {\it relative} 
ages will be important in our analysis.
Finally, as a further check of our results, in 
section 5.2. we will discuss a totally independent 
way of estimating the ages of the PN central stars
and its effect on the time variation of the
O/H abundance gradient.

%________________________________________________________________

\section{Results and discussion}

   \subsection{Time variation of the O/H gradient}

The PN sample has been divided into three groups
of increasing ages, namely Groups I, II, and III.
Since all objects have ages under
10 Gyr, we have considered initially
the following groups, which we will
call Case A: 
Group I:  $0 < \tau {\rm (Gyr)} < 3$,
Group II:  $3 < \tau {\rm (Gyr)} < 6$,
and Group III:  $ \tau  > 6\ {\rm Gyr}$. 

   \begin{figure*}
   \centering
   \includegraphics[width=13cm]{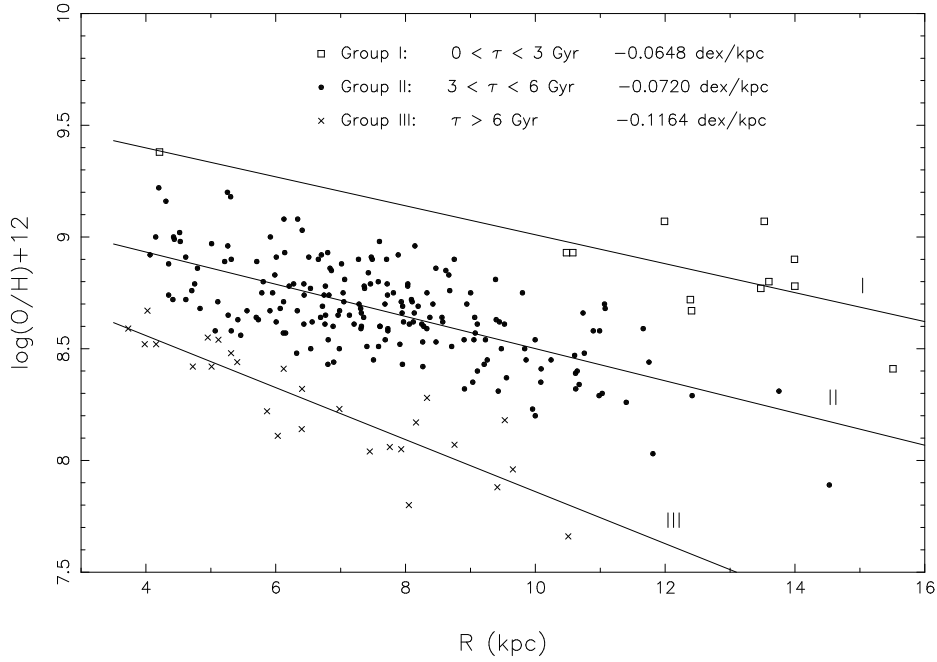}
      \caption{Radial O/H gradients from PN of Group I
      (squares), Group II (dots) and Group III
      (crosses) for Case A.
              }
         \label{fig4}
   \end{figure*}

   \begin{figure*}
   \centering
   \includegraphics[width=13cm]{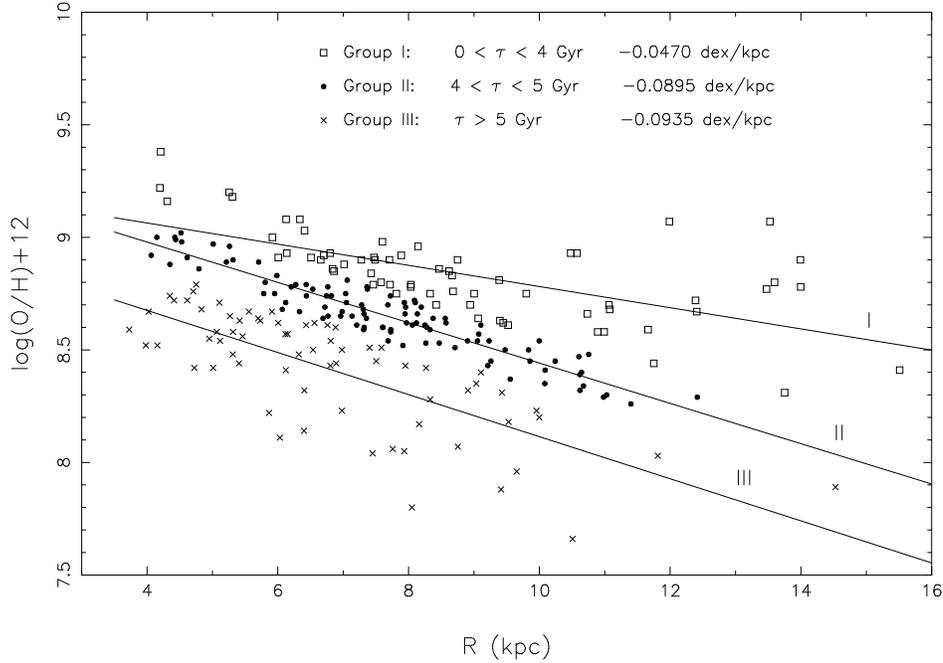}
      \caption{The same as Fig.~\ref{fig4} for Case B.
              }
         \label{fig5}
   \end{figure*}

The derived gradients are shown in Fig.~\ref{fig4}, 
where we have taken $R_0 = 8\,$kpc for the 
galactocentric distance of the LSR, in agreement 
with the value adopted by Edvardsson et al. 
(\cite{edvardsson}). In fact, the main results 
of this paper are unchanged if we adopt different 
values such as the IAU recommended value 
$R_0 = 8.5\,$kpc, or some recently determined 
result such as $R_0 = 7.6\,$kpc (see for example 
Maciel \& Quireza \cite{mq}). The figure 
shows the separation of Group I (squares), 
II (dots) and III (crosses), and the linear 
gradients of each group (dex/kpc). The values of the 
gradients (dex/kpc), the associated uncertainties 
$\sigma_g$, the correlation coefficients $r$
and the number of objects in each class are given in 
Table~1. We can see that Group~I is underpopulated 
compared to the remaining groups, reflecting the 
lack of the most massive and presumably younger 
objects, which affects the derived slope.
Therefore, we have alternatively defined Case B as
follows: Group I:  $0 < \tau {\rm (Gyr)} < 4$; 
Group II:  $4 < \tau {\rm (Gyr)} < 5$, and
Group III:  $\tau > 5\ {\rm Gyr}$, 
so that we have approximately the same number
of nebulae in each age group. The corresponding
results are shown in Fig.~\ref{fig5} and also in 
Table~1.

%%%%%%%%%%%%%%%%%%%%%%%%%%%%%%%%%%%%%%%%%%%%%%%%%%%%%%%%
  \begin{table}
     \caption[]{Results of the linear fits}
         \label{table1}
      $$
          \begin{array}{cccc}
            \hline
            \noalign{\smallskip}
            {\rm Case\ A}  & {\rm I} & {\rm II} & {\rm III} \\
            \noalign{\smallskip}
            \hline
            \noalign{\smallskip}
    {d\log{\rm (O/H)}\over dR} & -0.0648 & -0.0720 & -0.1164 \\
    \sigma_g   & 0.0168    &  0.0057   &  0.0127   \\
    r              & -0.77   & -0.67   & -0.88   \\
    n              & 12        & 195       &  27 \\
            \noalign{\smallskip}
            \hline
            \noalign{\smallskip}
            {\rm Case\ B}  & {\rm I} & {\rm II} & {\rm III} \cr
            \noalign{\smallskip}
            \hline
            \noalign{\smallskip}
    {d\log{\rm (O/H)}\over dR} & -0.0470 & -0.0895 & -0.0935 \\
    \sigma_g   & 0.0070    &  0.0034   &  0.0102   \\
    r              & -0.64   & -0.94   & -0.75   \\
    n              & 66        & 99        &  69 \\
            \noalign{\smallskip}
            \hline
         \end{array}
    $$  
   \end{table}
%%%%%%%%%%%%%%%%%%%%%%%%%%%%%%%%%%%%%%%%%%%%%%%%%%%%%%%%

From the figures and the uncertainties in the
slopes given in Table~1, it can be seen that there is a clear
tendency for the O/H gradient to flatten out with time.
This tendency is clear in both Cases A and B, and is
particularly strong between Groups III and II (Case A)
and Groups II and I (Case B). This is confirmed by 
the inspection of the uncertainties in the slopes 
as given in Table~1.  For Case A, 
the gradients have flattened from $-0.11\,$dex/kpc to
$-0.06\,$dex/kpc, while for Case B, we have
$-0.09\,$dex/kpc for the oldest group and
$-0.05\,$dex/kpc for the youngest one. Overall,
one could conclude that the gradients flattened
out from $-0.11\,$dex/kpc to $-0.06\,$dex/kpc in 
about 9 Gyr, or from $-0.08\,$dex/kpc to
$-0.06\,$dex/kpc in the last 5 Gyr only.

   \subsection{Ages of the central stars from their
    main sequence masses}

In view of the uncertainties involved in the estimate
of the ages of the stars in our sample, which are 
basically due to the the adopted age-metallicity 
relationship, it is interesting to evaluate the time 
variation of the abundance gradients based on 
independent age estimates. This can be achieved on 
the basis of a correlation between the N/O abundances 
and the central star masses recently discussed by 
Cazetta \& Maciel (\cite{cazetta}), which is supported 
by recent theoretical calculations (see for example Marigo
\cite{marigo}). According to this relation, if N/O 
is the nitrogen abundance relative to oxygen by number 
of atoms,  the core mass $M_c$ in solar  masses  is 
given by the relations

   \begin{equation}%eq.(4)
     M_c \simeq 0.7242 + 0.1742 \, \log({\rm N/O})
    \label{mcno1}
    \end{equation}

\noindent
for $\log({\rm N/O}) \leq -0.26$ and

   \begin{equation}%eq.(5)
     M_c \simeq 0.825 + 0.936 \, \log({\rm N/O}) + 
     1.439 \, [\log({\rm N/O})]^2
    \label{mcno2}
   \end{equation}

\noindent
for $\log({\rm N/O}) > -0.26$. These relationships are valid
in the approximate range $-1 < \log({\rm N/O}) < 0.2$.
The main sequence mass $M_{ms}$ can be obtained through a initial 
mass-final mass relation given by

   \begin{equation}%eq.(6)
     M_c \simeq 0.5 + 0.1 \, M_{ms} \ .
     \label{mcmms}
   \end{equation}

\noindent
This set of equations characterizes the so-called 
{\it low-mass calibration}  (see a detailed 
discussion in Cazetta \& Maciel 
\cite{cazetta} and Maciel \cite{maciel01}), and 
shows a better agreement with the recent 
discussions on the PN central star masses by 
Stasi\'nska et al. (\cite{stasinska}), Zhang 
(\cite{zhang}), with the mass-N/O abundance 
relation by Marigo et al. (\cite{marigo96}) and 
the initial mass-final mass relations by 
Groenewegen \& de Jong (\cite{groenewegen})
and Bl\"ocker \& Sch\"onberner (\cite{blocker}).

From Eqs.~\ref{mcno1}-\ref{mcmms} above the main sequence
mass can be estimated from the N/O abundances.
The stellar age can then be derived using 
the average lifetimes based on stellar 
evolutionary models for Population I stars
given by Bahcall \& Piran (\cite{bahcall}), 
that is

   \begin{equation}%eq.(7)
     \log t(M_{ms}) = 10.0 - 3.6\, \log M_{ms} 
     + (\log M_{ms})^2
     \label{tmms}
   \end{equation}

\noindent
where $t(M_{ms})$ is given in years and $M_{ms}$ 
again in solar masses. This relation has an estimated
uncertainty of about 10\% according to Bahcall
\& Piran (\cite{bahcall}) or somewhat larger 
according to Scalo (\cite{scalo}), but in any case 
it is expected to produce reliable {\it relative} 
ages.

We have applied the procedure described above
to the PN central stars in our sample, using
revised N/O abundances listed by Cazetta \&
Maciel (\cite{cazetta}) and Maciel \& Chiappini 
(\cite{maciel94}). Again dividing the objects
into Groups I, II and III, we obtained the
results shown in Fig.~\ref{fig6}, where the
symbols have the same meaning as in 
Fig.~\ref{fig4}. The slopes are given in the 
figure, and the corresponding uncertainties and 
correlation coefficients are in the range
$\sigma_g \simeq 0.013$ to 0.021 and 
$\vert r \vert \simeq 0.67$ to 0.88.
Although the available sample is smaller than 
in the case of the previous figures, 
and some superposition can be observed  
in the different classes, it is clear that the same 
pattern is observed here, that is, the youngest 
objects display flatter gradients, as can be 
seen from the slopes given on the top of the figure.
Since the adopted procedures to estimate the
ages in the case of Fig.~\ref{fig4} and 
Fig.~\ref{fig6} are totally independent from each 
other, we have a further evidence 
that the {\it relative} behaviour of the
gradients is indeed reflected by the results
shown in Figs.~\ref{fig4} and \ref{fig5}. 

 \begin{figure*}
   \centering
   \includegraphics[width=13cm]{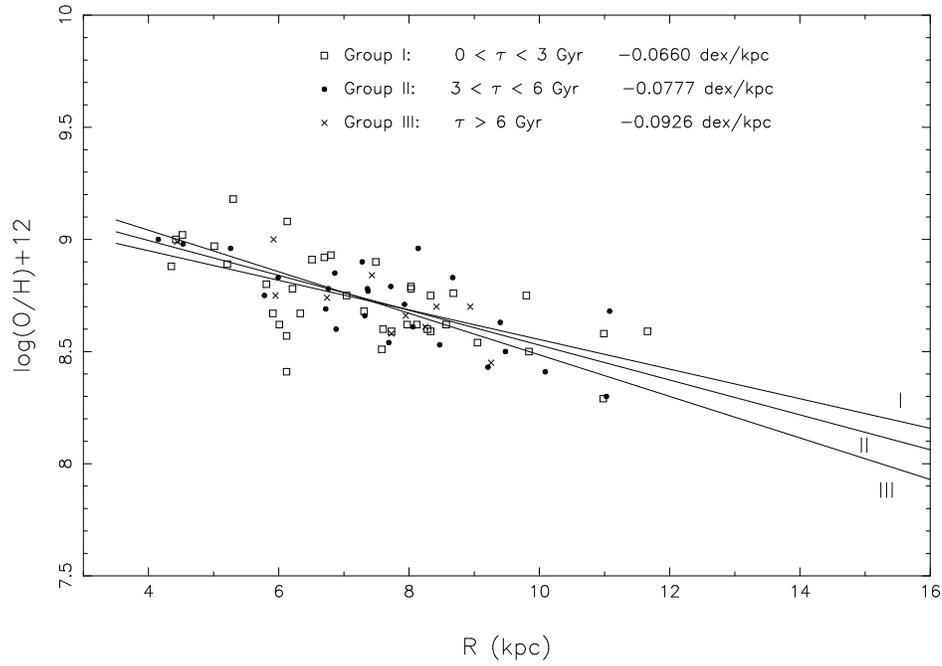}
      \caption{Radial O/H gradients from PN of Group I
      (squares), Group II (dots) and Group III
      (crosses) for Case A. The ages of the central stars
       have been calculated as described in Section 5.2.
              }
         \label{fig6}
   \end{figure*}

\subsection{Discussion}

A schematic plot of the results of Section 5.1. is shown
in Fig.~\ref{fig7}, where the solid lines refer
to Case A and B, and the dotted line is a reference 
line at $-0.07\,$dex/kpc. This reference line is 
usually considered as an average value for the younger 
population, comprising HII regions and B stars, but 
this should be regarded with caution in view of
the recent discussions on these objects as well as on 
open cluster stars and cepheids (see references in the 
Introduction). A realistic uncertainty for the slope of 
the young population can be roughly taken as 0.02 dex/kpc.
For comparison purposes, we also include in the figure the 
results of the theoretical models by Hou et al. (\cite{hou})
based on an inside-out scenario for the 
formation of the disk with metallicity dependent 
yields (dashed line). The agreement 
is remarkable, especially during the last 5
Gyr, which include most of our objects, and 
for which the derived ages are relatively 
more accurate. From these data, we can estimate 
an average flattening rate
of $0.002\,$dex kpc$^{-1}$ Gyr$^{-1}$ (Case A)
or a rate of about $0.004\,$dex kpc$^{-1}$ 
Gyr$^{-1}$, considering both cases in the last
5 Gyr, suggesting that the O/H gradient 
has not changed more than 30\% in average 
during the last few Gyr. At earlier times, 
however, our results are consistent with a 
steeper rate, although the corresponding 
uncertainties are larger. 
 
\begin{figure*}
   \centering
   \includegraphics[angle=-90,width=13cm]{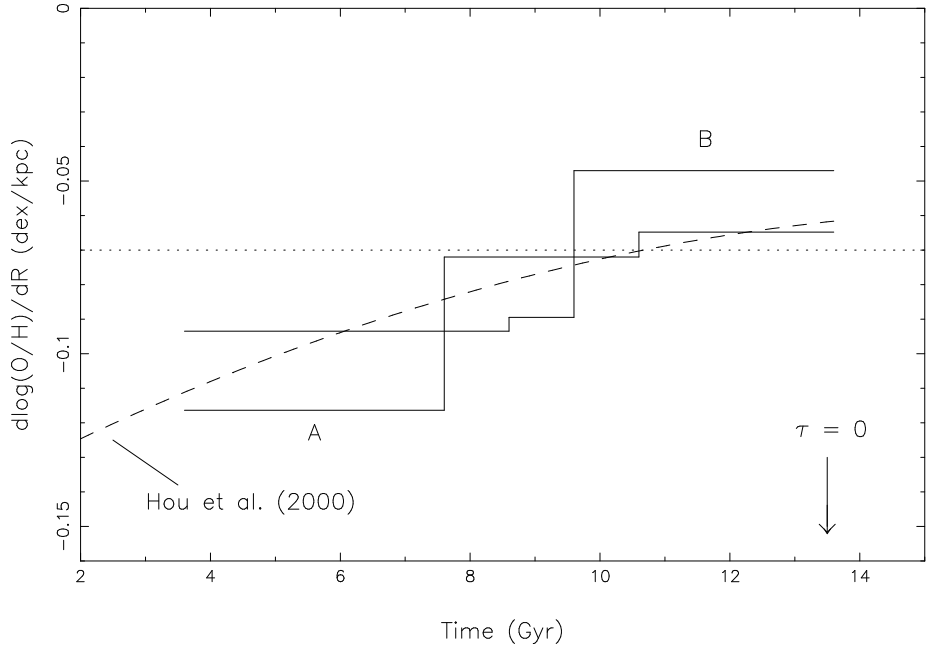}
      \caption{Time variation of the PN gradients
      for Case A and B (solid lines). Also shown 
      are results from theoretical models
      by Hou et al. (2000, dashed curve).
              }
         \label{fig7}
   \end{figure*}

It is interesting to compare these results with
earlier estimates based on planetary nebulae.
Maciel \& K\"oppen (\cite{mk}) found some evidences
for a steepening of the gradients, especially
those of Ne, Ar and S, while for the O/H ratio
some steepening could be obtained for PN of
Peimbert types II and III only. However, the average
ages attributed to the different PN types were
very approximate, and no effort was made to establish
individual ages. These average values have also
been used by Maciel \&	Quireza (\cite{mq}),
leading to the conclusion that the gradients 
steepen out in time at a ratio of about 0.004 
dex kpc$^{-1}$ Gyr$^{-1}$ in the Galaxy, 
a result which is not confirmed by the present paper.
The main difference between the present work and 
these early attempts, apart from our larger and
more accurate sample, refers to the definition
of the age groups. Maciel and K\"oppen (\cite{mk}) 
adopted literally the Peimbert classification  
scheme, which, strictly speaking, attributes 
different and increasing ages for the progenitor 
stars of PN of types~I, II and III, respectively.
In this work, we have not assumed any correlation 
between the PN type and the progenitor age, and 
made an effort to derive individual ages, so that we 
expect the present results to be more reliable. We 
notice that most Type~I PN in our sample belong 
to age Group~II, most Type~II PN are spread between 
Groups~I and II, while PN of Type~III belong either 
to Groups~II or III. Finally, most of the objects 
in Group~III are indeed Type~III PN (Case A). We can 
then conclude that within the disk PN of a given type 
there are objects with a reasonable age span. In  
other words, there seems to be some overlapping in the 
progenitor masses -- and ages -- within the Peimbert types, 
which was not accounted for by the earlier attempts
to estimate the time variation of the gradients,
and probably explains the different results presented
here as compared with the earlier work. In fact, 
as recently discussed by Peimbert \& Carigi (\cite{pc98}),
some overlapping occurs in the masses of the
progenitors of PN of types II and III, which 
supports our present views. 

\begin{acknowledgements}
      This work was partially supported by CNPq and FAPESP.
\end{acknowledgements}

%\newpage

\end{document}